# Persistent Spin Texture and Spin-Orbital Hall Responses on the AgI (110) Surface

Manish Kumar Mohanta
Department of Physics, Indian Institute of Technology Bhubaneswar, Bhubaneswar 752050, Odisha, India
manishkmr484@gmail.com, mkmohanta@iitbbs.ac.in

**Abstract:** A systematic investigation of the structural, electronic, and spin-orbital transport properties of the AgI (110) surface is presented using first-principles calculations combined with analytical modelling. The non-centrosymmetric and nonsymmorphic nature of the system gives rise to a robust persistent spin texture (PST), characterized by a unidirectional spin configuration and suppressed spin relaxation, enabling an effectively infinite spin lifetime. Unlike previously reported PST materials, which are predominantly based on chalcogen compounds, this work demonstrates that a halide semiconductor can host PST, thereby significantly expanding the materials platform for spintronic applications. The underlying mechanism is captured using an effective spin-orbit coupled Hamiltonian, which reproduces the anisotropic spin splitting and momentum shift observed in the band structure. This work introduces two new analytical models describing PST and compares them with existing models, offering new perspectives on PST arising from spin-orbit interaction. In addition, the system exhibits sizable intrinsic spin Hall conductivity (SHC) and orbital Hall conductivity (OHC), highlighting its potential for efficient charge-to-spin and charge-to-orbital conversion. The PST is found to be robust against biaxial strain, structural distortion, and multilayer formation, while a vertical electric field breaks the symmetry protection and drives a transition to a Rashba-type spin texture. These findings establish AgI (110) as a promising platform for realizing long-lived spin transport and tunable spin-orbit functionalities in the low-dimensional halide systems.

## 1. Introduction

Silver iodide (AgI) is a prototypical halide semiconductor that has attracted sustained interest due to its rich polymorphism and diverse functional properties spanning ionic conduction, optoelectronics, and phase-change behaviour. Depending on the temperature and pressure, AgI stabilizes in multiple crystalline phases, most notably the low-temperature $\beta$-phase (hexagonal wurtzite-type), the intermediate $\gamma$- phase (cubic zinc-blende or sphalerite structure), and the high-temperature $\alpha$- phase, which exhibits superionic conductivity. Among these, the $\gamma$-AgI phase, crystallizing in the cubic zinc-blende structure with space group $F\bar{4}3m$, is of particular interest due to its well-defined tetrahedral coordination, where each Ag atom is surrounded by four I atoms and vice versa. This highly symmetric structure serves as a fundamental platform for exploring electronic and symmetry-driven phenomena in halide systems. [1–6]

The (110) surface of zinc-blende crystals is especially significant, as it represents the natural cleavage plane characterized by an arrangement of equal numbers of cations and anions in a zigzag chain configuration. This geometry not only minimizes surface energy but also preserves key symmetry elements, including nonsymmorphic operations, which can give rise to unconventional electronic and spin textures. While extensive studies on zinc-blende semiconductors such as CdTe and ZnTe have revealed intriguing spin-orbit coupling effects and surface states, halide-based counterparts remain comparatively unexplored in the context of spin-dependent phenomena. Recent advances in spintronics have highlighted the importance of symmetry-protected spin textures, particularly the persistent spin texture (PST), which enables suppression of spin relaxation and effectively infinite spin lifetime under ideal conditions. [7,8] To date, PST has been predominantly reported in chalcogenide-based systems, where the interplay between crystal symmetry and spin-orbit coupling leads to unidirectional spin configurations. [9–14] However, extending this concept to halide semiconductors is both fundamentally appealing and technologically relevant, as these materials often exhibit tunable electronic properties, relatively weak spin-orbit coupling strength, and enhanced chemical flexibility.

In this context, the AgI (110) surface in the zinc-blende phase offers a promising yet largely unexplored platform for realizing symmetry-driven spin phenomena. Its nonsymmorphic symmetry, coupled with the reduced dimensionality of the surface, provides favorable conditions for stabilizing anisotropic spin splitting and novel spin textures.

Furthermore, understanding the interplay between crystal structure, spin-orbit coupling, and external perturbations such as strain and electric fields is essential for designing next-generation spintronic and orbitronics devices based on halide materials.

## 2. Computational Details

The ab initio calculations were performed using the Quantum Espresso package. [15,16] Fully relativistic norm-conserving pseudopotentials from the pseudoDojo [17] library was employed to accurately account for spin-orbit coupling. A k-point mesh of 15 ×15 ×1 was used in the self-consistent calculations. The convergence criteria for total energy and forces were set to $1.0\times10^{-6}$ Ry and $1.0\times10^{-3}$ Ry/Bohr respectively. Plane-wave kinetic energy cutoff and charge density cutoff values were fixed at 80 Ry and 500 Ry. Electronic occupations were treated using Gaussian smearing with a smearing width of 0.01 Ry. A vacuum space of more than 20Å is added to avoid periodic interaction along the *z*-direction. A set of maximally localized Wannier functions [18,19] was generated using the Wannier90 code. [20,21] Subsequently the spin Hall conductivity and orbital Hall conductivity were computed via the Kubo formula using a dense 150×150×1 *k*-mesh:

$$\sigma_{xy}^{z} = \frac{e}{\hbar}\int_{Bz}\frac{dk}{(2\pi)^2}\Omega^{z}(k)$$

$\Omega^z(k)$ is the k-resolved term which is given by $\Omega^z(k) = \sum_n f_{kn}\Omega_n^z(k)$. Here, $f_{kn}$ is the Fermi-Dirac distribution function for the *n*th band at k and $\Omega_n^z(k)$ is an analogue of the Berry curvature for the *n*th band given as;

$$\Omega_n^z(k) = \sum_{n'\neq n}\frac{2Im[\langle kn|j_x^z|kn'\rangle\langle kn'|v_y|kn\rangle]}{(\epsilon_{kn}-\epsilon_{kn'})^2}$$

Here, $j_x^z = \frac{1}{2}\{s_z, v\}$ is the spin current operator, $s_z = \frac{\hbar}{2}\sigma^z$ is the spin operator, $v$ is the velocity operator and $|kn\rangle$ is the wave function of energy $\epsilon_{kn}$.

Other packages, such as PYPROCAR and WannierTools, were used for postprocessing. [22,23] The ferroelectric properties are calculated using the Berry phase method. [24–26] The phonon dispersion was obtained using the Phonopy code. [27]

## 3. Results and Discussions

**3.1 Crystal geometry and stability:** The (110) facet of cubic zinc blende AgI belongs to the nonsymmorphic space group No. 31 with group order four, incorporating fundamental symmetry elements such as rotation, mirror and translational operations. The crystal remains invariant under the following symmetry operations: (i) the identity $\langle E|\{0,0,0\}\rangle$; (ii) a two-fold rotation about the $x$-axis combined with a fractional translation, $\langle C_{2x}|\{\frac{a}{2},\frac{b}{2},0\}\rangle$; (iii) a mirror reflection with respect to the zx-plane, $\langle M_{zx}|\{0,0,0\}\rangle$; and (iv) a glide reflection $\bar{M}_{xy}$ $\langle M_{xy}|\{\frac{a}{2},\frac{b}{2},0\}\rangle$, where $a$ and $b$ denote the in-plane lattice constants. The top and side views of the primitive structural unit of the (110) facet are illustrated in Figure 1(a). The unit cell contains four atoms, giving rise to a total of twelve phonon modes, comprising three acoustic and nine optical branches, as shown in Figure 1(b). The absence of imaginary frequencies in the phonon spectrum confirms the dynamical stability of the structure.

**3.2 Crystal symmetry and ferroelectric polarization:** The (110) AgI crystal structure is non-centrosymmetric, as evident from the top view geometry, and exhibits a pronounced in-plane ferroelectric polarization of 1.2 nC/m. For comparison, related materials such as $\beta$-GeSe(~ 0.16 nC/m) [28] and SnTe (~ 0.194 nC/m) [29] show significantly smaller in-plane polarization. The emergence of ferroelectricity in this system can be attributed to the broken inversion symmetry in conjunction with the underlying translational symmetry of the lattice.

**3.3 Electronic properties of the basic building block of (110) AgI:** The electronic band structure, shown in Figure 1(c), reveals a direct bandgap of 1.85 eV with both the conduction band minimum (CBM) and the valence band maximum (VBM) located at the Brillouin zone centre (Γ-point). The dispersions of the CBM and VBM in the vicinity of Γ exhibit anisotropic behaviour, as further corroborated by the two-dimensional projection of the conduction and valence bands presented in Figure 1(d) and Figure S1 in Supplemental Material (SM). The carrier effective masses, extracted from the band curvature, are summarized in Table S1. In addition, the anisotropy in effective mass has been quantified by evaluating the ratio along the $\Gamma \rightarrow X$ to $\Gamma \rightarrow Y$ directions. The resulting ratios indicate a pronounced anisotropy, being approximately six for holes and two for electrons, consistent with the trends discussed above. The orbital-resolved band structure, shown in Figure 1(e), reveals that the conduction band is predominantly derived from Ag: *s* states with a noticeable contribution from I: *d* orbitals, whereas the VBM is mainly composed of I: *p*, I: *d,* and Ag: *d* states. The band decomposed charge density also indicates that the CBM is largely localized on Ag atoms, with a minor contribution from I atoms, while the VBM is primarily distributed over the I sublattice. Notably, the presence of *d*-orbital character in both the CBM and VBM suggests significant orbital hybridization, which can play a crucial role in determining the electronic response and transport properties. Additional orbital-projected band structures are plotted in Figure S2.

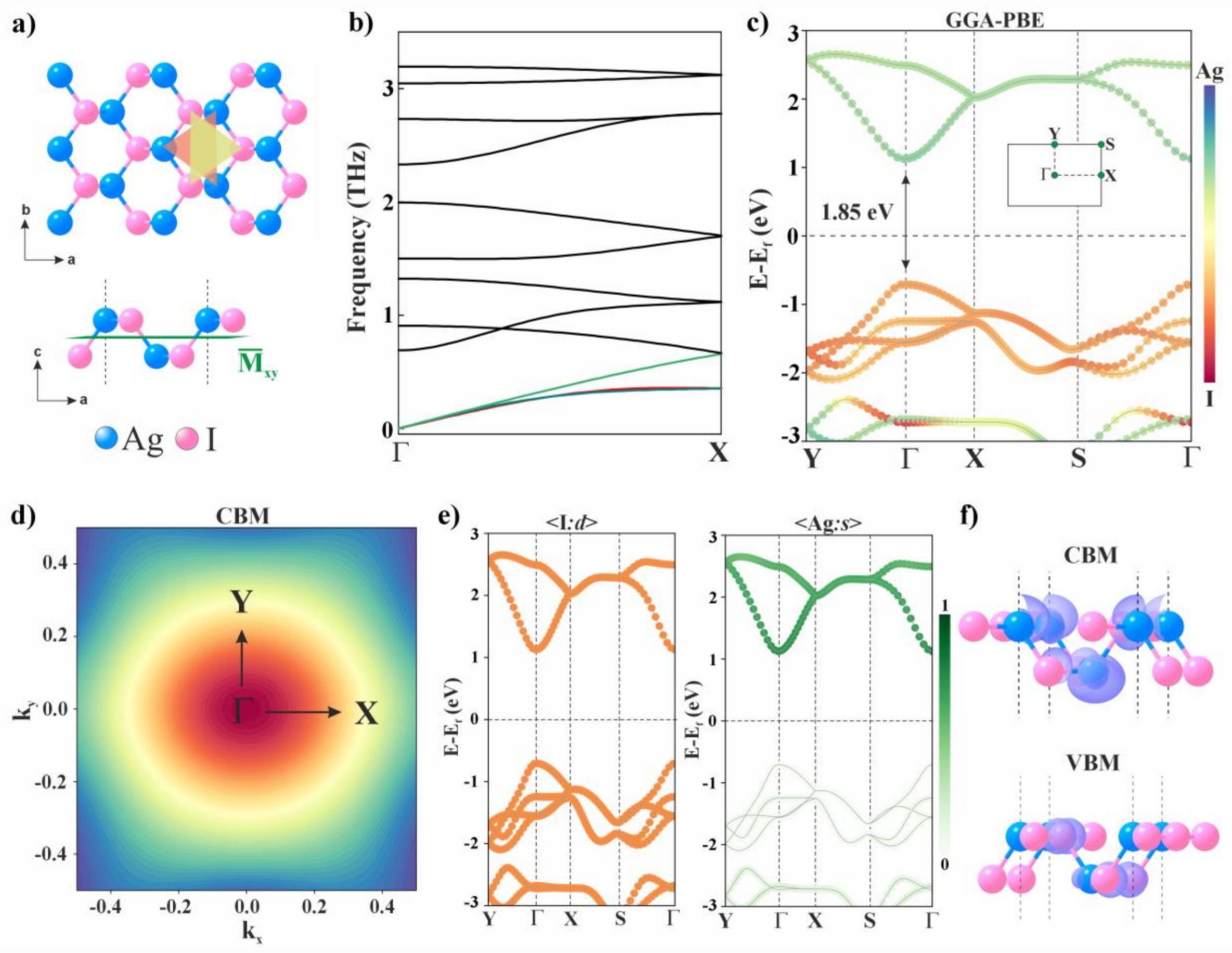

Figure 1: (a) Top and side views of Ag (110) surface, (b) phonon dispersion spectrum confirming the dynamical stability of the structure, (c) atom-projected electronic band structure, (d) two-dimensional projection of CBM highlighting its dispersion near the Γ-point, (e) orbital-resolved band structure, (f) band decomposed charge density corresponding to the CBM and VBM.

**3.4 Spin-orbit coupling and analytical model of AgI (110):** In this section, relativistic effects are incorporated into the electronic structure via spin-orbit coupling (SOC), and the resulting spin-projected band structure is presented in Figure 2(a). A characteristic band splitting associated with the persistent spin texture (PST) is observed around the Γ-point: the spin degeneracy is lifted along the $\Gamma \rightarrow Y$ direction, while the bands remain degenerate along $\Gamma \rightarrow X$, reflecting a strong directional anisotropy spin splitting. Such a type of spin splitting is a key signature of PST. Furthermore, the spin-projected Fermi surface, shown in Figure 2(b), clearly demonstrates the hallmark features of PST, in agreement with the theoretical predictions of Schliemann et al. [30] and Mohanta and Jena [9]. The spin-split bands are polarized along $\pm\hat{s}_z$ as shown in Figure S3, and form a persistent spin texture (PST) around the Brillouin zone centre. The momentum offset ($\vec{Q}$) between the spin-up and spin-down parabolic dispersions is clearly evident, as illustrated in Figure 2(b), reflecting the characteristic shift associated with PST.

The spin-orbit coupling constant for CBM and VBM is estimated using the relation [31–33] $\gamma = \frac{2\Delta E}{\Delta k}$, yielding the values of 0.3 eV.Å for CBM and 0.48 eV.Å for VBM. These values indicate relatively weak SOC compared to typical chalcogenide semiconductors such as CdTe (1.35 eV.Å), ZnTe (1.37 eV.Å), GeTe (1.67 eV.Å), while remaining comparable to materials like SnSe (0.74 eV.Å), GeSe (0.57 eV.Å). [10,12] Further calculations reveal that the material exhibits an intrinsic spin Hall conductivity (SHC) of approximately ~90 $(\hbar/e)$S/cm as shown in Figure 2(d), which is comparable to that reported for CdTe ~100 $(\hbar/e)$S/cm. [10] The $k$-point resolved SHC, presented in Figure 2(e), indicates that the dominant contributions originate from states in the vicinity of the Γ-point within the rectangular Brillouin zone.

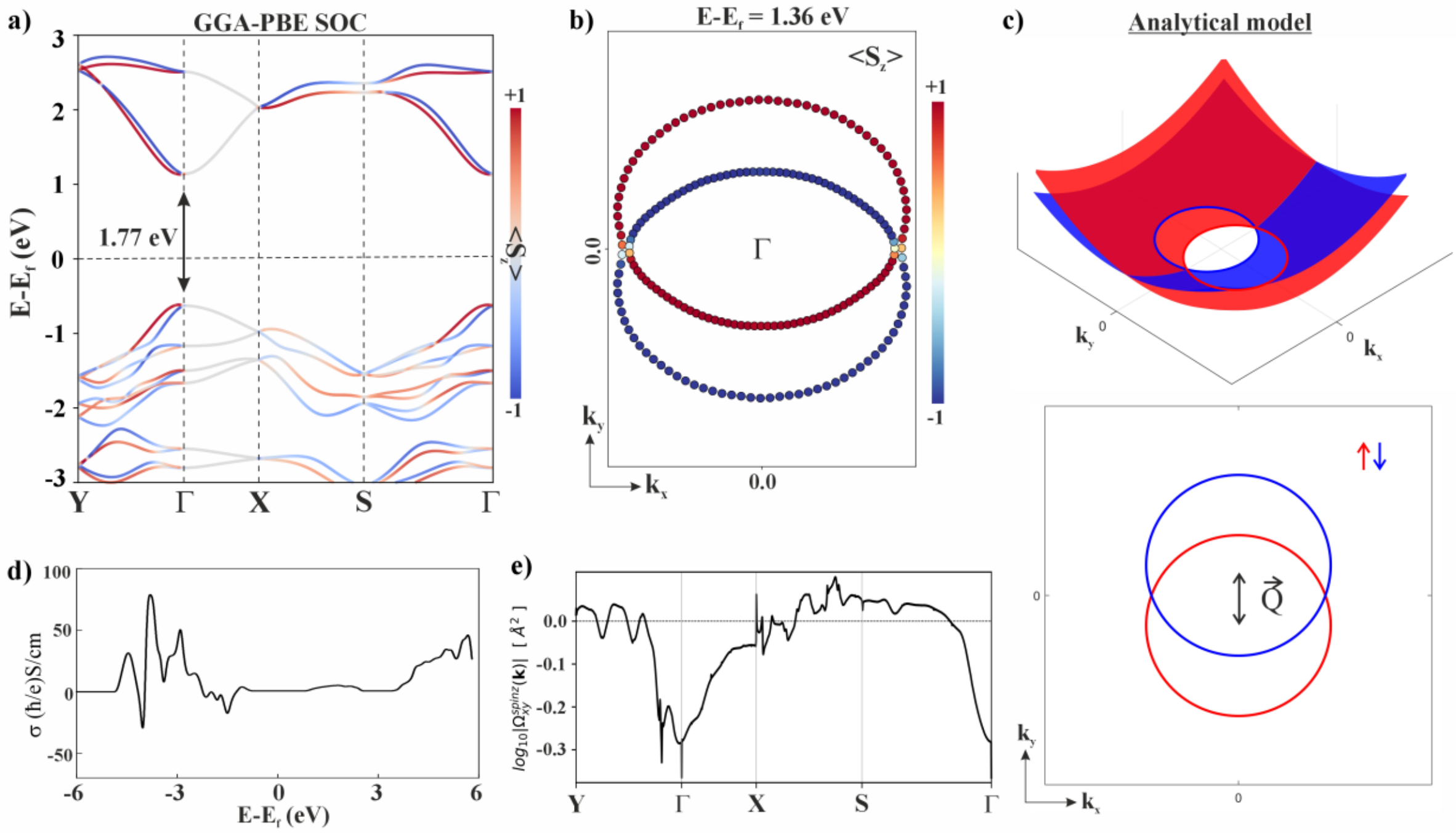

Figure 2: (a) Electronic band structure including relativistic effect, (b) spin-projected Fermi surface illustrating the characteristic feature of PST, (c) spin-projected band dispersion, and corresponding constant energy contour plots obtained from the model Hamiltonian [Eq. (5)], (d) spin Hall conductivity as a function of energy, (e) $k$-point resolved distribution of the SHC in the Brillouin zone.

### 3.5 Analytical Models representing PST:

To gain insight into the origin of such spin splitting, a mathematical model can be derived from the spin-orbit coupled Hamiltonian of an electron moving in an effective electric field, given by;

$$H_{SO} = \left[\frac{\hbar^2}{4m_0^2c^2}\right][\nabla V(r) \times p].\sigma \text{ ........... (1)}$$

where $V(r)$ denotes the crystal potential, $p$ is the electron momentum operator, $\sigma$ represents Pauli spin matrices. By recognizing that $\nabla V(r)$ acts as an effective electric field E, the above expression can be simplified to a compact form (Eq. 2),

$$H_{SO} = C[\mathrm{E} \times p].\sigma \text{ .............. (2)}$$

where $C$ is an effective spin-orbit coupling constant. To further elucidate the origin of the spin splitting, we construct a model Hamiltonian that explicitly depends on the direction of the applied electric field. Considering first a unit out-of-plane electric field oriented along the $\hat{z}$ direction, and restricting the system to two dimensions, Eq. (2) can be reduced to a simpler form. This leads Eq. (3), which corresponds to the well-known Rashba Hamiltonian [34].

$$H_{Rashba} = \alpha(k_x\sigma_y - k_y\sigma_x) \text{ ................ (3)}$$

Interestingly, if a unit in-plane electric field is applied along $\hat{x}$, and considering the same 2D case, equation (2) can be reduced to the equation of persistent spin texture (PST) [35] given by;

$$H_{PST} = \gamma k_y\sigma_z \text{ ............... (4)}$$

Here $\alpha, \gamma$ are SOC constants of corresponding SOC interactions. The energy eigenvalues and eigenstates ($\phi$) of Eq. (4) are $E_{1/2} = \pm\gamma k_y$ and $\phi_1 = \begin{pmatrix}1\\0\end{pmatrix}$, $\phi_2 = \begin{pmatrix}0\\1\end{pmatrix}$. The spin polarization of each energy band is given by $\langle\phi_i|\vec{\sigma}|\phi_i\rangle$, resulting in polarization vectors $\begin{pmatrix}0\\0\\1\end{pmatrix}$ and $\begin{pmatrix}0\\0\\-1\end{pmatrix}$.

The total Hamiltonian incorporating the SOC term relevant to the PST can be written as;

$$\mathcal{H} = \frac{\hbar^2}{2m}\left(k_x^2 + k_y^2\right) + \gamma k_y\sigma_z \text{ ............. (5)}$$

where $\gamma$ characterizes the strength of the spin-orbit interaction. The spin-projected energy dispersion derived from this model is presented in Figure 2(c). A comparison between the constant-energy contours obtained from first-principles (DFT) calculations and those from the analytical model reveals a clear qualitative agreement, confirming that the essential features of the persistent spin texture are well captured by the effective Hamiltonian.

In addition, previous studies have proposed independent theoretical frameworks to describe persistent spin texture (PST), most notably by Schliemann et al. ($\mathcal{H}_S$) [30] and Mohanta & Jena ($\mathcal{H}_{MJ1}$) [35], which provides deeper insight into its underlying physics. The corresponding model Hamiltonians are given as follows:

$$\mathcal{H}_S = \mathcal{H}_{Rashba} + \mathcal{H}_{Dresselhaus-1} = \alpha(\sigma_x k_y - \sigma_y k_x) + \beta(\sigma_x k_x - \sigma_y k_y) \text{ ...... (6)}$$

$$\mathcal{H}_{MJ1} = \mathcal{H}_{Rashba} + \mathcal{H}_{Dresselhaus-2} = \alpha(\sigma_x k_y - \sigma_y k_x) + \mathcal{M}(\sigma_x k_y + \sigma_y k_x) \text{ ...... (7)}$$

$$\mathcal{H}_{MJ1} \equiv \alpha(\sigma_z k_y - \sigma_y k_z) + \mathcal{M}(\sigma_z k_y + \sigma_y k_z) \text{ ...... (8)}$$

From these two Hamiltonians, the conditions required to realize a persistent spin texture (PST) are given by $\alpha = \pm\beta$ or $\alpha = \pm\mathcal{M}$. Among the two models, the Hamiltonian proposed by Mohanta & Jena ($\mathcal{H}_{MJ1}$) offers a more straightforward analytical framework compared to ($\mathcal{H}_S$), as Eq. (4) can be directly obtained from Eq. (8) under $\alpha = +\mathcal{M}$ condition. It is worth noting, however, that these conditions are rather stringent; even a slight deviation such as $\beta$ *or* $\mathcal{M}$ being approximately 99% of $\alpha$ can lead to a partial PST, as demonstrated by Mohanta and Jena in ref. [35].

Let's introduce two new analytical models proposed for PST, which incorporate combined Weyl and Dresselhaus spin-orbit interaction terms, expressed as follows;

$$\mathcal{H}_{MKM2} = \mathcal{H}_{Weyl} + \mathcal{H}_{Dresselhaus-1} = \mathcal{J}(\sigma_x k_x + \sigma_y k_y) + \beta(\sigma_x k_x - \sigma_y k_y) \text{ ......... (9)}$$

$$\mathcal{H}_{MKM3} = \mathcal{H}_{Weyl} + \mathcal{H}_{Dresselhaus-2} = \mathcal{J}(\sigma_x k_x + \sigma_y k_y) + \mathcal{M}(\sigma_x k_y + \sigma_y k_x) \text{ ........ (10)}$$

The conditions for PST obtained from Eq. (9-10), arise when the Weyl coupling equals either the Dresselhaus-1 or Dresselhaus-2 term, i.e., $\mathcal{J} = \pm\beta/\mathcal{M}$. For example, under the condition $\mathcal{J} = -\beta$, the Hamiltonian $\mathcal{H}_{MKM2}$ simplifies to $\mathcal{H}_{MKM2} = 2\mathcal{J}\sigma_y k_y$ which represents a PST state. Including the free-electron energy contribution, the Hamiltonian for the $\mathcal{H}_{MKM2}$ model is written as: $\mathcal{H}_{MKM2} = \frac{\hbar^2}{2m}(k_x^2 + k_y^2) + \mathcal{J}(\sigma_x k_x + \sigma_y k_y) + \beta(\sigma_x k_x - \sigma_y k_y)$. The corresponding energy eigenvalues, eigenstates, and spin polarization under the condition of $\mathcal{J} = -\beta$, are summarized in Table-1. The spin textures for each band, obtained under different combinations of $\mathcal{J}$ and $\beta$, are presented in Figure 3.

**Table 1: The energy eigenvalues, corresponding eigenstates, and spin polarization of the Hamiltonian $\mathcal{H}_{MKM2}$**

| Cases | Energy eigenvalues | Eigenstates | Spin polarization |
|---|---|---|---|
| $\mathcal{H}_{MKM2} = 2\mathcal{J}\sigma_y k_y$ | $E_3 = \frac{\hbar^2}{2m}(k_x^2 + k_y^2) + 2\mathcal{J}k_y$ | $\begin{pmatrix} -i \\ 1 \end{pmatrix}$ | $\begin{pmatrix} 0 \\ -1 \\ 0 \end{pmatrix}$ |
| | $E_4 = \frac{\hbar^2}{2m}(k_x^2 + k_y^2) - 2\mathcal{J}k_y$ | $\begin{pmatrix} i \\ 1 \end{pmatrix}$ | $\begin{pmatrix} 0 \\ 1 \\ 0 \end{pmatrix}$ |

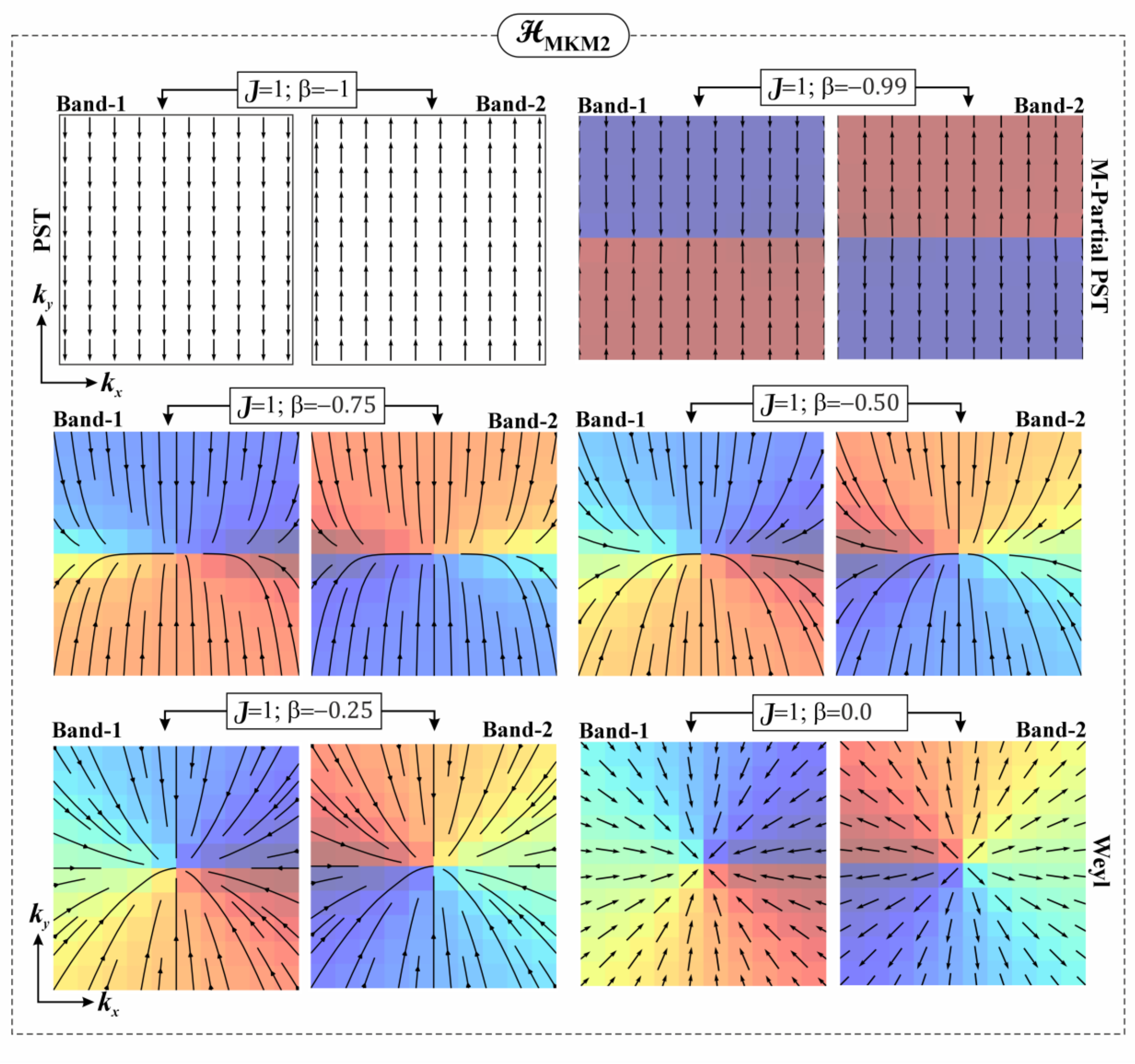

Figure 3: The spin textures obtained from the Hamiltonian $\mathcal{H}_{MKM2}$ (Eq. 9) for different values of SOC constants. For the plots, the momentum-space range $[k_x, k_y]$ was chosen as [-1,1].

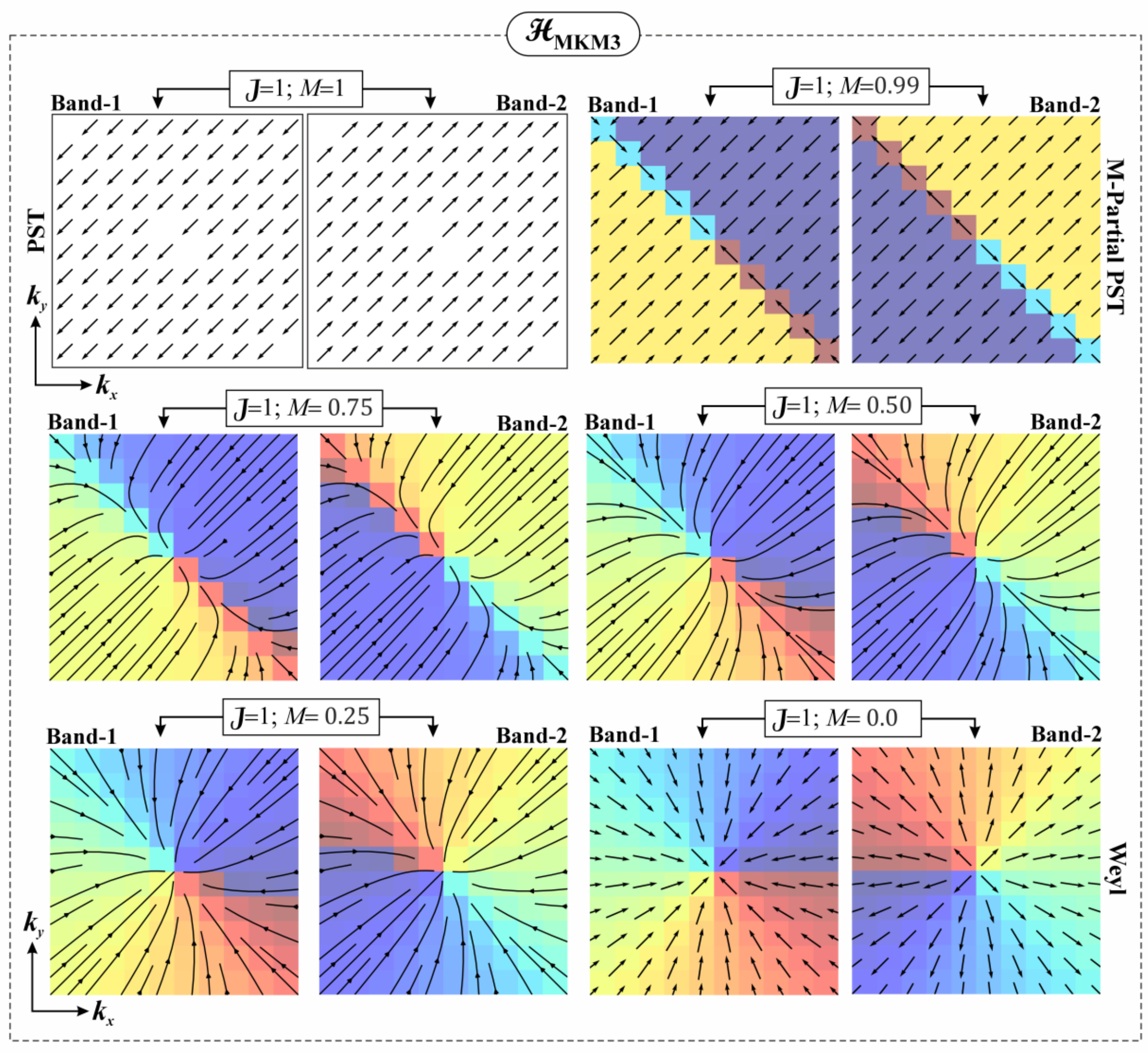

Figure 4: The spin textures obtained from the Hamiltonian $\mathcal{H}_{MKM3}$ (Eq. 10) for different values of SOC constants. For the plots, the momentum-space range $[k_x, k_y]$ was chosen as [-1,1].

The energy eigenvalues obtained from $\mathcal{H}_{MKM3} = \mathcal{J}(\sigma_x k_x + \sigma_y k_y) + \mathcal{M}(\sigma_x k_y + \sigma_y k_x)$ under the condition of $\mathcal{J} = \mathcal{M}$ are given by; $E_{5/6} = \mp\sqrt{2}\mathcal{J}(k_x + k_y)$. The corresponding spinors are $\frac{1}{\sqrt{2}}\begin{pmatrix} 1 \\ \mp e^{i\frac{\pi}{4}} \end{pmatrix}$. The spin textures for each band, obtained under different combinations of $\mathcal{J}$ and $\mathcal{M}$ are presented in Figure 4. The spin-projected Fermi surfaces obtained from $\mathcal{H}_{MKM3}$, including the free electron energy term, are shown in Figure 5. For comparison, the spin-projected Fermi surfaces derived from Eqs. (6-7) are also included in Figure 5, highlighting the similarities and differences among the various models. It is interesting to note that a new type of partial PST, termed M-partial PST, has been identified in comparison with the cases obtained from Eqs. (6-7). This distinction becomes evident when compared with earlier reports [14,35].

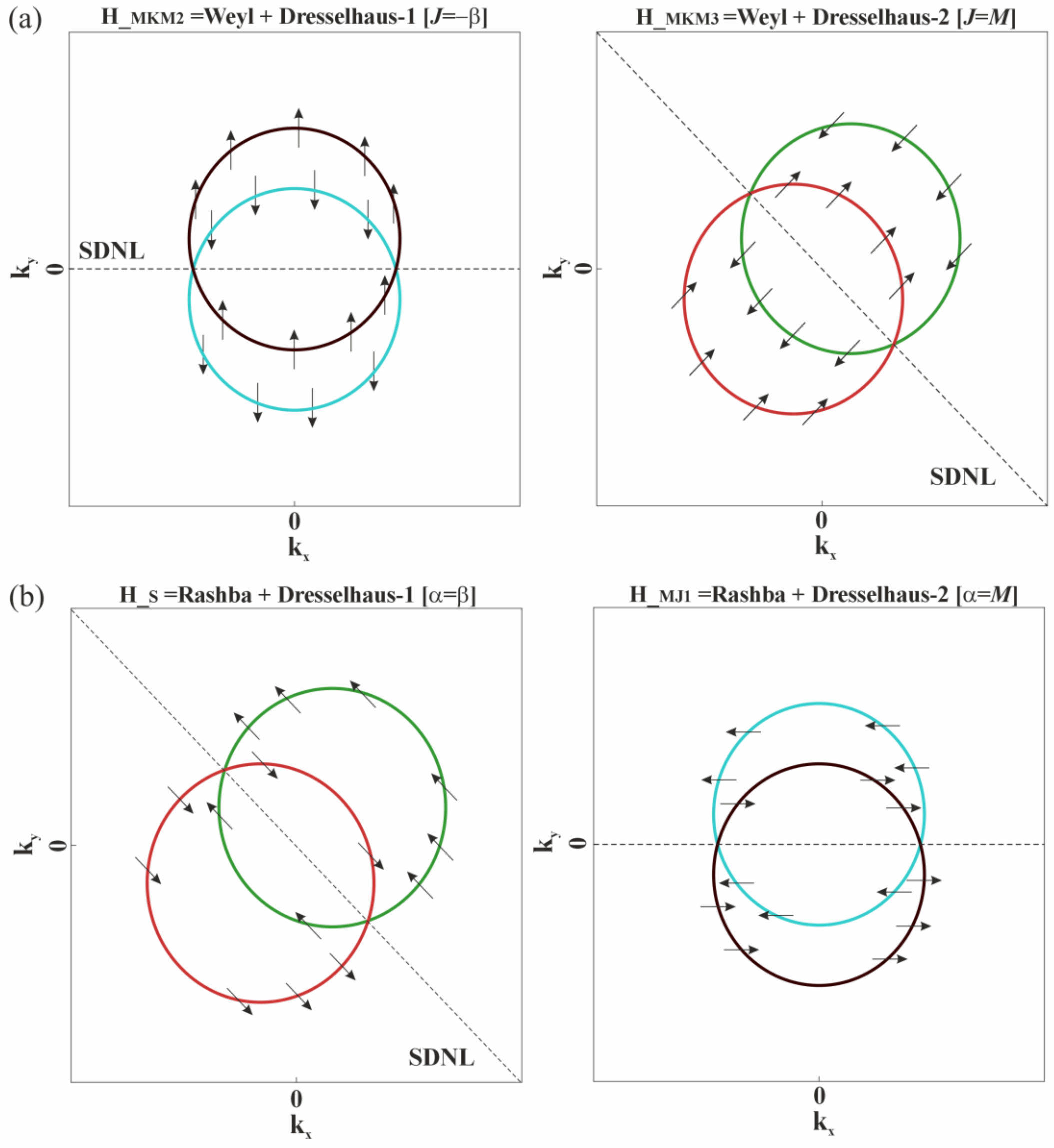

Figure 5: A comparative spin-projected Fermi surface plot obtained from different analytical models proposed in (a) this work, and (b) previous reports.

**3.6 Effect of strain and vertical electric field on PST:** External perturbations such as mechanical strain or an applied electric field can significantly modify the electronic properties of a material by altering its crystal symmetry, atomic spacing, and potential landscape. Strain changes the interatomic distances and orbital overlap, which can lead to variations in band dispersion, band gap, and carrier effective mass. Depending on the magnitude and direction, strain may induce band gap tuning and semiconductor-metal transitions. Similarly, an external electric field perturbs the electrostatic potential within the material, leading to band bending, charge distribution, and possibly lifting of band degeneracies. In low-dimensional systems, electric fields can also break inversion symmetry and induce phenomena such as Rashba-type spin splitting or polarization switching. Therefore, external perturbations provide an effective route for tuning the electronic structure and transport properties of materials, offering important opportunities for designing functional electronic and optoelectronic devices.

Biaxial tensile and compressive strains up to $\pm 4\%$ are applied, as schematically illustrated in the inset, and the corresponding electronic band structures are presented in Figure 6. The results indicate a systematic reduction in the band gap under strain. Notably, compressive strain is more effective than tensile strain in modulating the band gap of the AgI (110) surface. Despite these changes, the characteristic features of the PST remain robust under biaxial strain. The spin-projected band structures further confirm that the spin polarization ($\pm \hat{s}_z$) of the bands near the Fermi level is preserved, indicating that the underlying spin texture is largely unaffected by strain.

The spin Hall effect (SHE) and orbital Hall effect (OHE) are transverse transport phenomena where an applied electric field E generates a perpendicular flow of angular momentum without net charge current; in the SHE, this manifests as a spin current, whereas in OHE, it corresponds to a flow of orbital angular momentum. Microscopically, both can be described within linear response theory using the Kubo formula, where the Hall conductivities are given by Berry curvature-like terms in momentum space: the spin Hall conductivity (SHC) is $\sigma_{xy}^{s} = \frac{e}{\hbar} \sum_n \int \frac{d^3k}{(2\pi)^3} f_n(k) \Omega_{n,xy}^{s}(k)$, and the orbital Hall conductivity (OHC) is $\sigma_{xy}^{L} = \frac{e}{\hbar} \sum_n \int \frac{d^3k}{(2\pi)^3} f_n(k) \Omega_{n,xy}^{L}(k)$, where $\Omega^{s}$ and $\Omega^{L}$ are spin- and orbital- Berry curvatures derived from interband matrix elements of spin $\hat{s}_z$ and orbital angular momentum $\hat{L}_z$, respectively. While SHE typically relies on SOC, OHE can arise even without SOC due to orbital texture in Bloch states, though SOC often converts orbital currents into spin currents, linking OHC and SHC. The effect of biaxial strain on spin Hall conductivity and intrinsic orbital Hall conductivity can be seen in Figure 7, which indicates a significant enhancement of both SHC and OHC under compressive strain. In materials exhibiting both SHC and OHC, the interplay allows efficient charge to spin and charge to orbital conversion, which is crucial for spin-orbit torque devices, non-volatile memory, and emerging orbitronics, where orbital currents can enhance or even dominate spin transport efficiency in low-SOC systems.

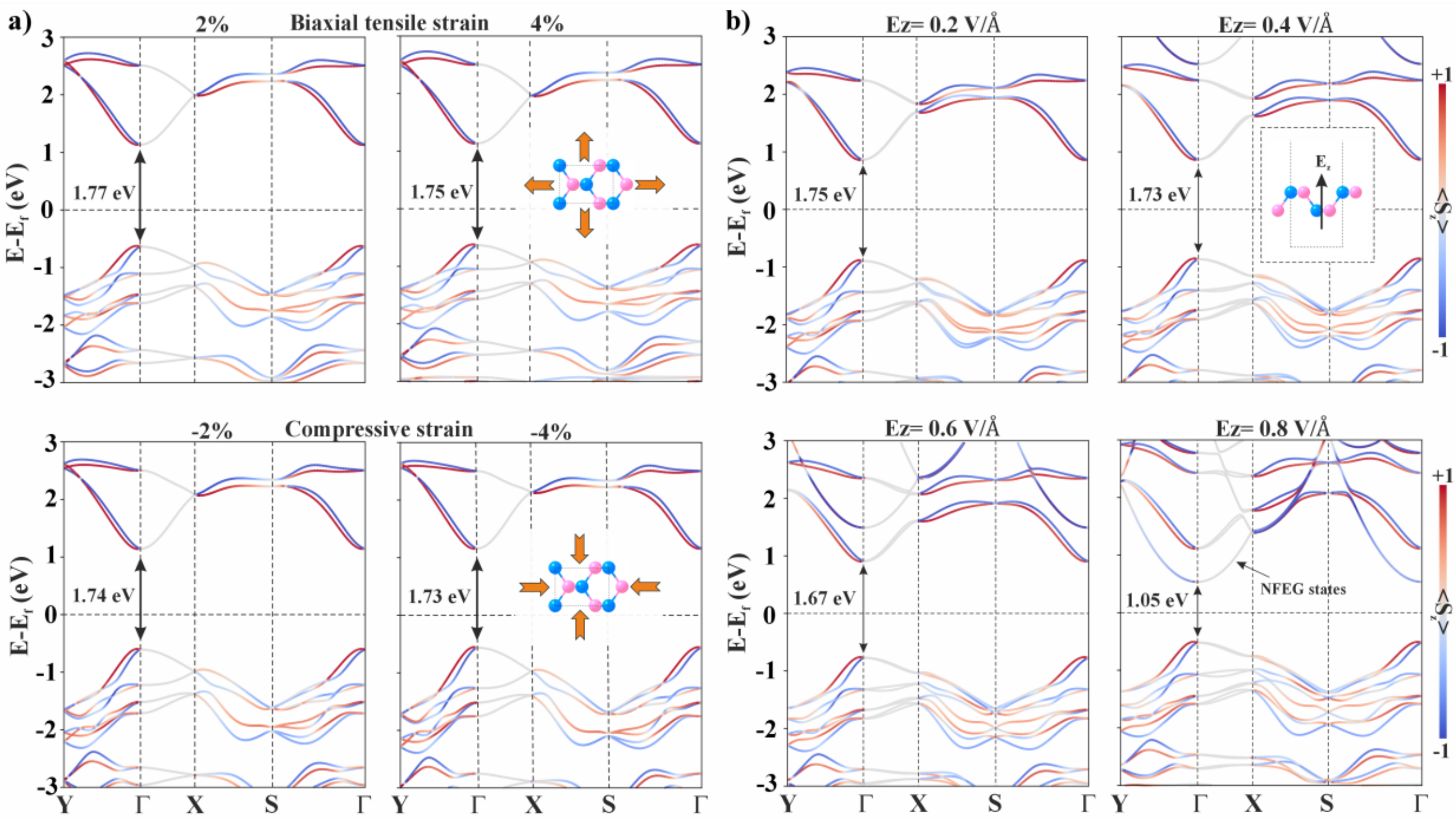

Figure 6: Modulation of the electronic band structure under external perturbations: (a) biaxial strain, and (b) applied vertical electric field.

A vertical electric field of increasing strength is applied to the AgI (110) surface, and the corresponding field-modulated band structures are shown in Figure 6(b). The results reveal a progressive reduction in the band gap with increasing electric field strength. In addition, the bands along the $\Gamma \rightarrow X$ directions, which were previously degenerate, become nondegenerate under the applied field, exhibiting a finite spin splitting. A comparative analysis highlighting this effect is provided in Figure S4 for clarity. Notably, the application of the vertical electric field breaks the persistent spin texture, indicating that the symmetry conditions required to sustain PST are no longer preserved. Figure S5 further illustrates that, at an applied electric field strength of $E_z$=0.8 V/Å, nearly free electron gas (NFEG)-like states emerge as surface-localized bands with parabolic dispersion approaching the Fermi level. This figure also shows that the electronic states near the Fermi level acquire finite in-plane spin components ($\hat{s}_x$ and $\hat{s}_y$). This is further confirmed by the spin-projected Fermi surfaces plotted in Figure 8; for the pristine and biaxially strained cases, the spin texture is purely out-of-plane ($\hat{s}_z$), whereas under a vertical electric field, the same bands exhibit mixed spin character with finite $\hat{s}_x, \hat{s}_y$ and $\hat{s}_z$ components. The application of a vertical electric field drives the system from a PST phase to a non-PST phase, effectively transitioning toward a Rashba-type spin texture characterized by finite in-plane spin components ($\hat{s}_x$ and $\hat{s}_y$). Additionally, the plane-averaged electrostatic potential along the z-direction under different perturbations is presented in Figure S6, highlighting the field-induced asymmetry responsible for this transition.

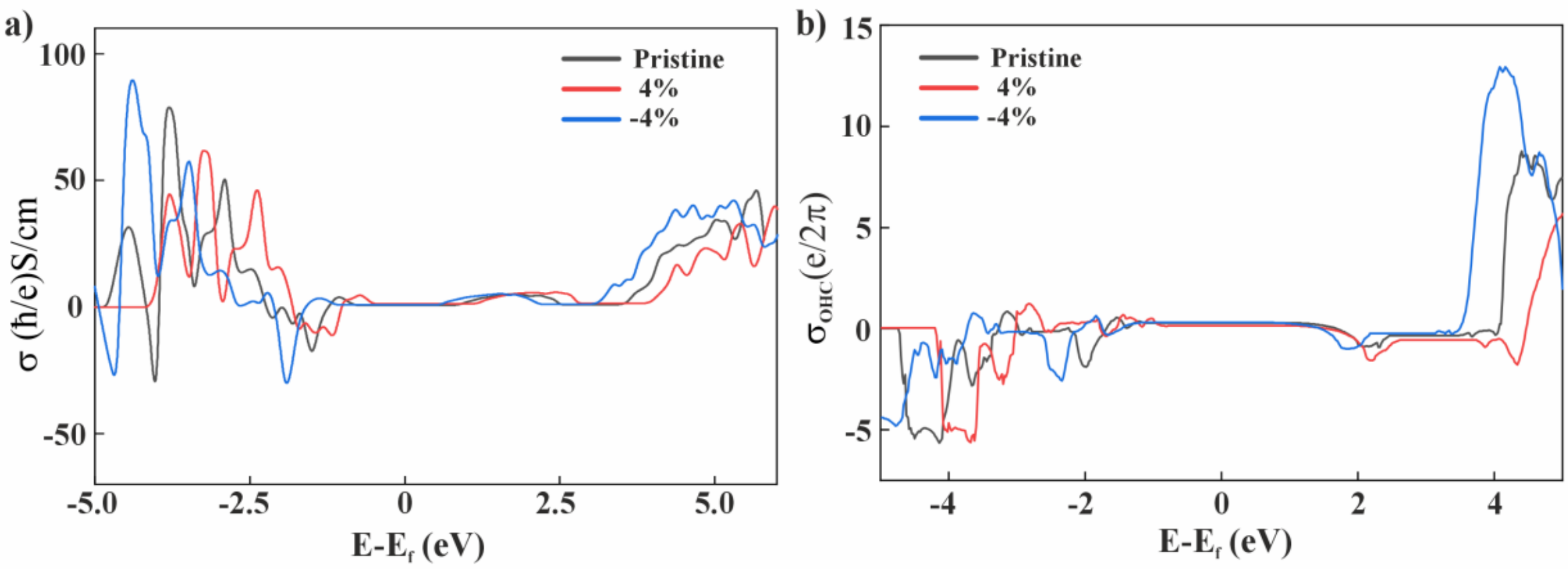

Figure 7: (a) Spin Hall conductivity and (b) orbital Hall conductivity as a function of energy.

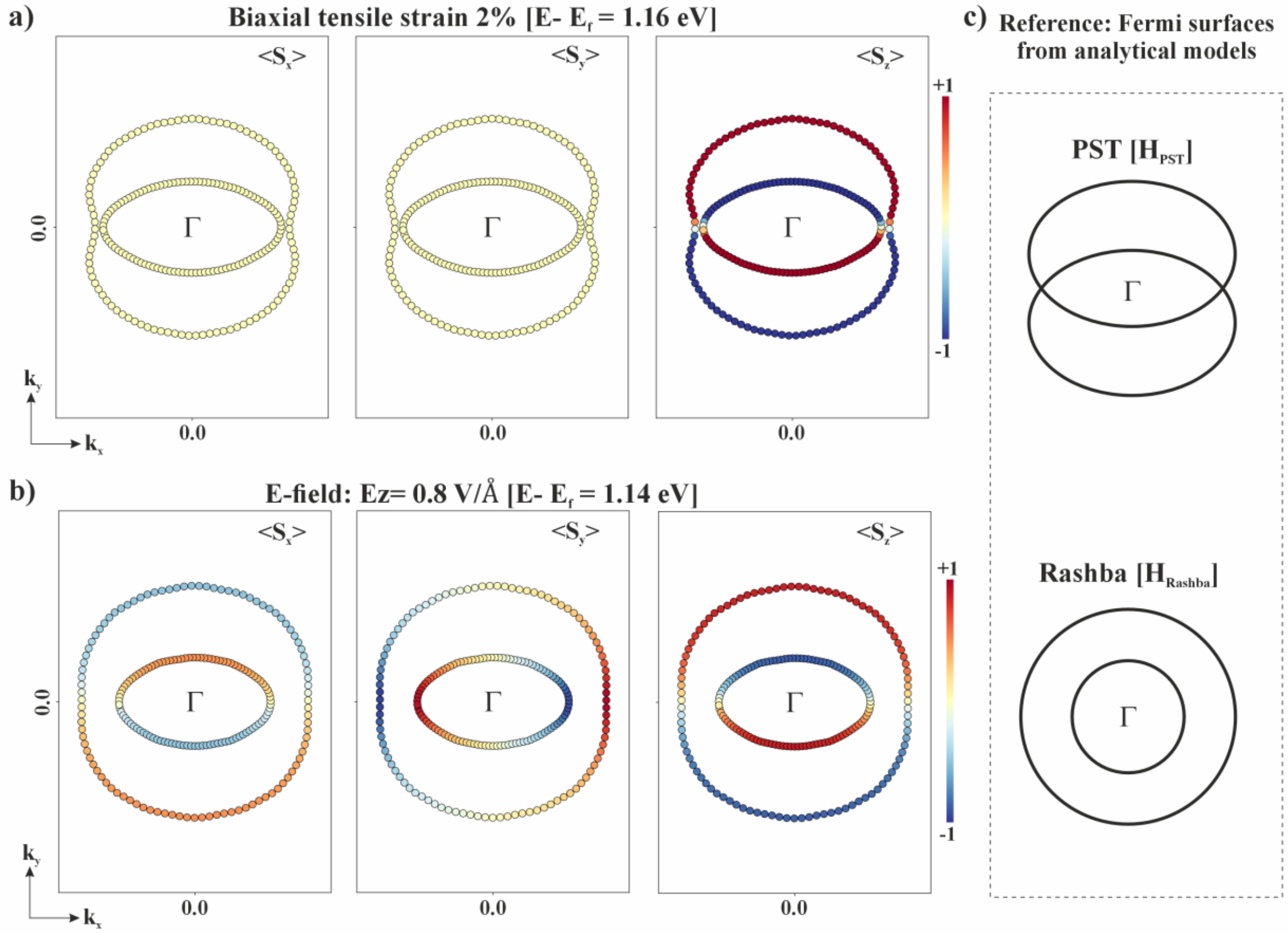

Figure 8: Spin-projected Fermi surfaces under (a) biaxial strain, (b) applied vertical electric field, (c) reference Fermi surfaces were obtained from the analytical models for comparison.

**3.7 Electronic properties of multilayer AgI:** To explore the experimental realization of PST, a multilayer (4L) AgI (110) is constructed. The crystal structure retains its nonsymmorphic character along the growth direction. Within the basic structural unit, the coordination environment of $Ag^{+}$ ions differs between surface and interior: surface Ag atoms are threefold coordinated, whereas those in the inner layers exhibit fourfold coordination. As a consequence, the surface atoms undergo structural relaxation and distortion to minimize the surface energy. Notably, similar systems, such as ZnSe, exhibit a strong preference for the (110) orientation and remain stable at room temperature [36], supporting the feasibility of such configurations. The geometric structures of the modelled slabs are shown in Figure 9, where surface distortions are evident. Despite these structural modifications, the spin-projected band structures and constant energy Fermi surface confirm the preservation of PST, demonstrating its robustness against both biaxial strain and structural distortion. Figure 9(d-e) reveals a substantial enhancement in both the SHC and OHC for a multilayer AgI (110) surface.

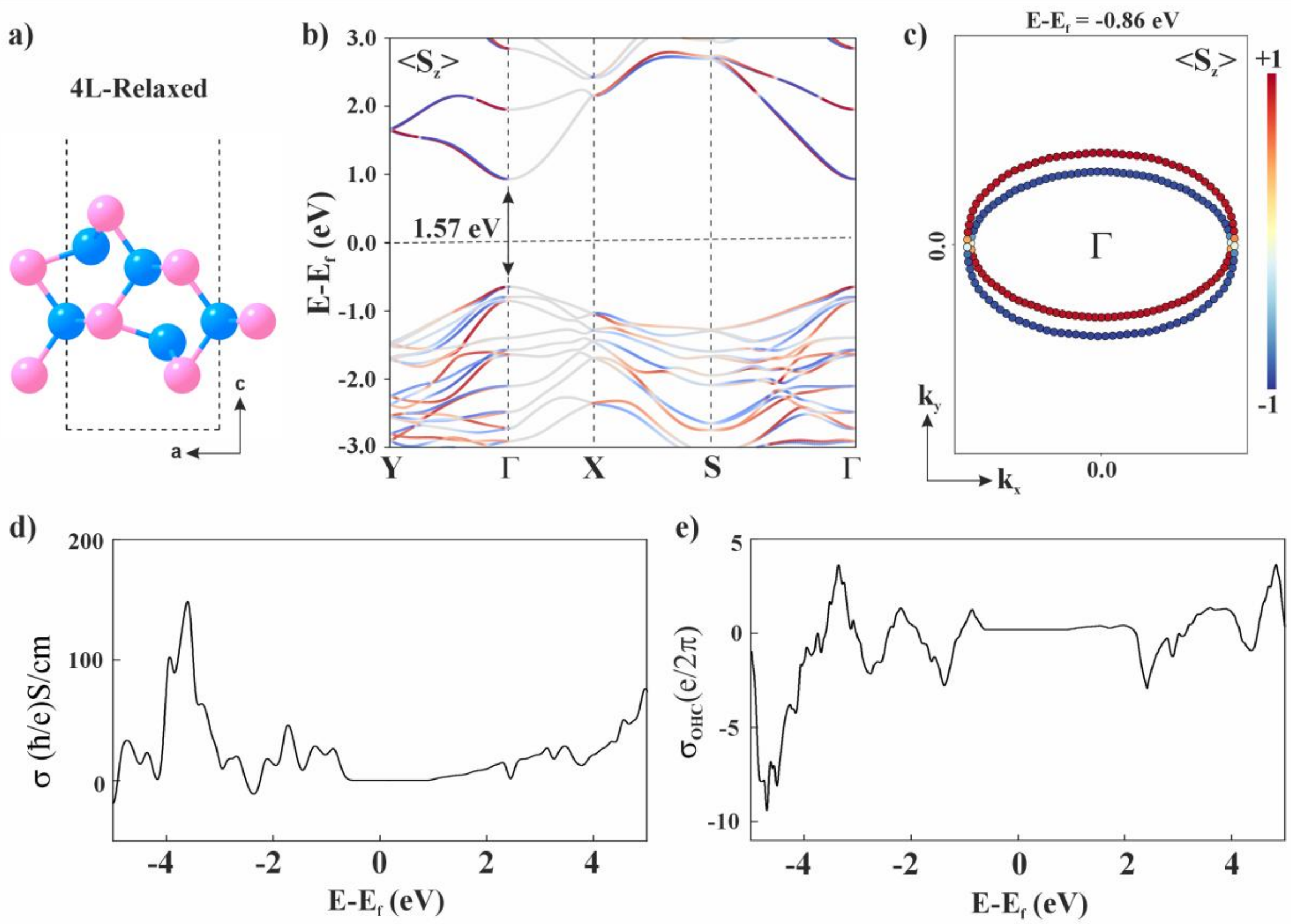

Figure 9: (a) Geometrical view of the relaxed multilayer AgI (110) structure, (b) corresponding spin-projected band structure, (c) spin-projected Fermi surface, (d) spin Hall conductivity, and (e) orbital Hall conductivity.

## 4. Conclusion

In summary, a systematic investigation of the structural, electronic, and spin-orbital transport properties of the AgI (110) surface is carried out using first-principles calculations and analytical modelling. The system is found to host a robust persistent spin texture (PST) arising from its nonsymmorphic symmetry, leading to highly anisotropic spin splitting with preserved spin polarization near the Fermi level. An effective model Hamiltonian is developed to capture the essential features of the spin-orbit coupling, showing good agreement with the first-principles results. In addition to PST, the material exhibits appreciable intrinsic spin Hall conductivity and orbital Hall conductivity, highlighting its potential application for spintronic and orbitronics applications. The PST and associated spin characteristics are shown to be remarkably robust against biaxial strain and structural distortions, including in multilayer configurations, indicating favourable prospects for experimental realization. However, the application of a vertical electric field breaks the symmetry protection, driving a transition from a PST to a Rashba-type spin texture with mixed spin components. Furthermore, an enhanced SHC and OHC in multilayer structures suggest the possibility of tuning spin-orbital responses via thickness engineering. Overall, this work establishes AgI (110) as a promising platform for realizing controllable spin and orbital transport phenomena in low-dimensional systems.

**Acknowledgements:** M. K. Mohanta acknowledges financial support from Anusandhan National Research Foundation (ANRF) under the Ramanujan Faculty Fellowship (Grant No. RJF/2025/000601). The author gratefully acknowledges the research infrastructure and facilities provided by the Indian Institute of Technology-Bhubaneswar. MKM acknowledges National Supercomputing Mission (NSM) for providing computing resources of 'PARAM RUDRA' at IIT-Patna, which is implemented by C-DAC and supported by the Ministry of Electronics and Information Technology (MeitY) and Department of Science and Technology (DST), Government of India.